\documentclass[11pt,twoside,a4paper]{article}
\usepackage{epsfig}
\usepackage{graphicx}
\usepackage{srt_style}
\begin{document}
\baselineskip .5cm
%
\title{Observations of extragalactic H$_{2}$O masers in bright IRAS sources}

\author{A. Tarchi$^{1}$, C. Henkel$^{1}$, A. B. Peck$^{1}$, L. Moscadelli$^{2}$ and K. M. Menten$^{1}$}

\affil{$^{1}$ Max-Planck-Institut f\"ur Radioastronomie, Auf dem H\"ugel 69, D-53121 Bonn, Germany\\
$^{2}$ Osservatorio Astronomico di Cagliari, Loc.\ Poggio dei Pini, Strada 54, 09012 Capoterra (CA), Italy}

%
\begin{abstract}
We report the first results of an ongoing survey at 22\,GHz with the 100-m Effelsberg telescope to search for water maser emission in bright IRAS sources. We have detected water vapor emission in IC~342. The maser, associated with a star forming region $\sim$10--15\arcsec\ west of the nucleus, consists of a single 0.5\,km\,s$^{-1}$ wide feature and reaches an isotropic luminosity of 10$^{-2}$\,L$_{\odot}$ ($D$=1.8\,Mpc). Our detection raises the detection rate among northern galaxies with IRAS point source fluxes $S_{\rm 100\mu m}$ $>$ 50\,Jy to 16\%.
\end{abstract}

\section{Introduction}
To date, luminous extragalactic H$_2$O masers can be grouped into three classes: those tracing accretion disks (e.g.\ NGC~4258); those in which the emission is either the result of an interaction between the radio jet and an encroaching molecular cloud or an accidental overlap, along the line-of-sight, between a warm dense molecular cloud and the radio continuum of the jet (e.g.\ NGC~1052 and Mrk~348); those related to prominent sites of star formation, such as the ones observed in M~33 (the earliest known extragalactic H$_2$O masers), and later also in IC~10.

Extragalactic H$_2$O masers are preferentially detected in nearby galaxies that are bright in the infrared (\cite{braatz97}). While nuclear masers are of obvious interest, non-nuclear masers are also important for a number of reasons: these sources allow us to pinpoint sites of massive star formation, to measure the velocity vectors of these regions through VLBI proper motion studies, and to determine true distances through complementary measurements of proper motion and radial velocity (e.g. \cite{greenhill93}). 

We have therefore observed the nearby spiral galaxies IC~342 and Maffei~2, both of which exhibit prominent nuclear bars and strong molecular, infrared, and radio continuum emission. In the following we report the results of our observations.

\section{Water maser observations of Maffei~2 and IC~342}

\subsection{The observation and image processing}
The $6_{16} - 5_{23}$ line of H$_2$O (rest frequency: 22.23508\,GHz) was observed with the 100-m telescope of the MPIfR at Effelsberg\footnote{The 100-m telescope at Effelsberg is operated by the Max-Planck-Institut f\"ur Radioastronomie (MPIfR) on behalf of the Max-Planck-Gesellschaft (MPG).} toward Maffei~2 and IC~342. The full width to half power beamwidth was 40\arcsec. The observations were carried out in a dual beam switching mode with a beam throw of 2\arcmin\ and a switching frequency of $\sim$1\,Hz. The pointing accuracy was always better than 10\arcsec.

On May 12, 2001, IC~342 was observed with the Very Large Array\footnote{The National Radio Astronomy Observatory is a facility of the National Science Foundation operated under cooperative agreement by Associated Universities, Inc.} (VLA) in its B configuration. The 103 channels used, out of the 128 observed, cover a range in velocity of $\sim$8\,km\,s$^{-1}$ centered at 16\,km\,s$^{-1}$ LSR (the velocity of the line detected at Effelsberg). No continuum subtraction was needed. The data were deconvolved using the CLEAN algorithm (\cite{hoegbom74}). The restoring beam is $\rm 0\farcs4 \times 0\farcs3$ and the rms noise per channel is $\sim$10\,mJy, consistent with the expected thermal noise.

\subsection{Results}

Our single-dish observations towards Maffei~2 yielded no detection, with a 5$\sigma$ upper limit of 25\,mJy (channel spacing: 1.05\,km\,s$^{-1}$; velocity range: --250\,km\,s$^{-1}$ $<$ $V$ $<$ +230\,km\,s$^{-1}$; epoch: Apr 3, 2001; position: $\alpha_{2000}$ = 02$^{\rm h}$ 41$^{\rm m}$ 55$\fs$2, $\delta_{2000}$ = +59$^{\circ}$ 36$\arcmin$ 11$\arcsec$).

\begin{figure}
\centering
\includegraphics[width=8cm]{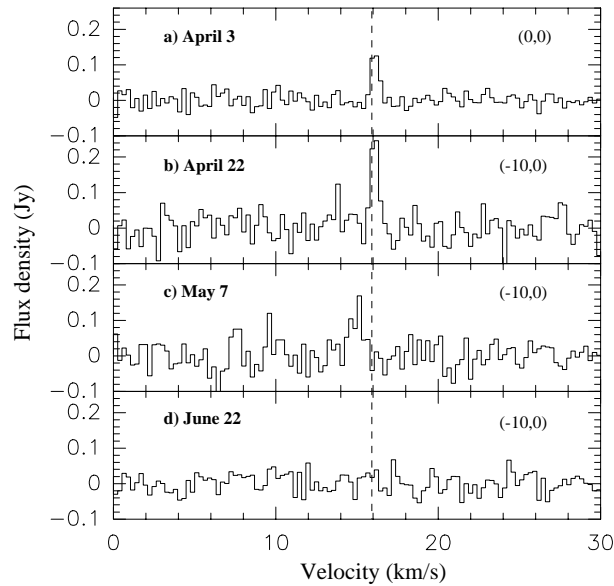}
\caption{The H$_2$O maser feature observed with high velocity resolution (channel spacings after smoothing four contiguous channels are 0.26\,km\,s$^{-1}$) on {\bf {a)}} April 3, {\bf {b)}} April 22, {\bf {c)}} May 7, and {\bf {d)}} June 22. The first spectrum has been taken at the (0\arcsec,0\arcsec) position, the others at (--10\arcsec,0\arcsec) relative to the position given in footnote `a' of Table 1. The dashed line indicates $V_{\rm LSR}$ = 16\,km\,s$^{-1}$. \label{flare}}
\end{figure}

\begin{figure}
\centering
\includegraphics[width=12cm]{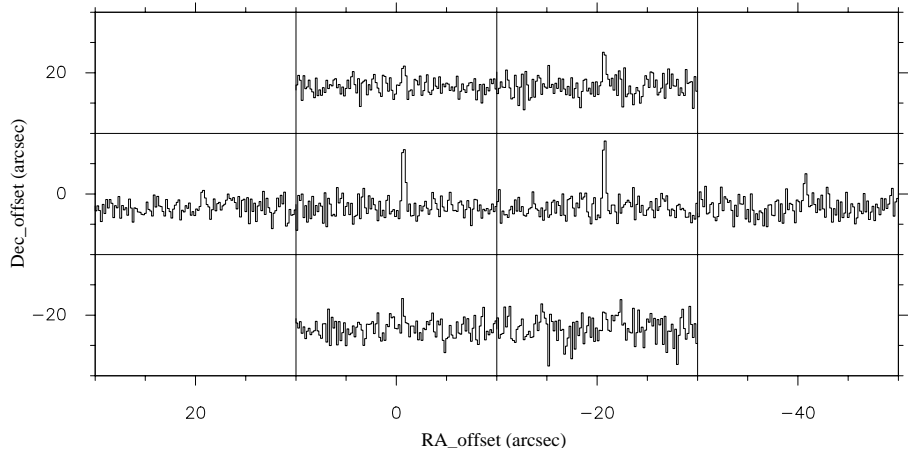}
\caption{H$_2$O spectra obtained toward the central region of IC~342. Positions are offsets relative to \mbox{$\rm \alpha_{2000} = 03^{h} 46^{m} 48\fs6$} and \mbox{$\delta_{2000} = +68^{\circ} 05\arcmin 46\arcsec$}. Note that the spacing between two individual spectra (20\arcsec) is approximately half of the size of the 100-m Effelsberg telescope beam at 22\,GHz (40\arcsec). Averaging four contiguous channels, the spectra have been smoothed to a spacing of 0.26~km\,s$^{-1}$. \label{map}}
\end{figure}

On April 2, 2001, we obtained the first definite detection of water vapor emission in IC~342. During the next night the detection was confirmed with a velocity resolution sufficient to resolve the line profile (Fig.\,\ref{flare}a). The detected feature lies at $V_{\rm LSR}$ = 16\,km\,s$^{-1}$ and has a linewidth of $\sim$0.5\,km\,s$^{-1}$; on April 2, no other component was seen at velocities --175\,km\,s$^{-1}$ $<$ $V$ $<$ +310\,km\,s$^{-1}$ (channel spacing: 1.05\,km\,s$^{-1}$; 5$\sigma$ noise level: 16\,mJy). A high line intensity ($\sim$100\,mJy) and good weather conditions allowed us to map the emitting region (Fig.\,\ref{map}). Further measurements were performed on April 22 and May 7 (Fig.\,\ref{flare}b\,\&\,\ref{flare}c). It is noticeable a fading in the peak and integrated intensities and a blue-shift in the velocity of the line by $\sim$1\,km\,s$^{-1}$. Unfortunately, no emission above $\sim$30\,mJy (3$\sigma$ level; 0.08\,km\,s$^{-1}$ channel spacing) was detected in the 22~GHz VLA B-array data taken on May 12. This fading by at least a factor of 3 within 5 days implies a size scale of $\la$\,$1.5\times10^{16}$\,cm ($\la$\,900\,AU) or $\la$\,0.5\,mas at a distance of 1.8\,Mpc (\cite{mccall89}; see also Sect.\,4.3); the corresponding brightness temperature is $\ga$10$^{9}$\,K. The most recent spectrum was obtained on June 22 at Effelsberg (Fig.\,\ref{flare}d). Confirming the VLA result, no maser signal was seen above 30\,mJy (3$\sigma$; channel spacing: 1.05\,km\,s$^{-1}$).

Fitting a synthetic Gaussian to the data taken on April 3 (see Fig.\,\ref{map}), i.e. minimizing the sum of the difference squared between calculated and observed peak and integrated flux densities, we can obtain an accurate position of the emitting region: $\alpha_{2000}$ = 03$^{\rm h}$ 46$^{\rm m}$ 46\fs3, $\delta_{2000}$ = +68$^{\circ}$ 05\arcmin 46\arcsec. This is 13\arcsec\ to the west of our center position that coincides with the optical nucleus and the 2$\mu$m peak \cite{kruit73}; \cite{becklin80}. The accuracy of the derived maser position has been estimated to be $\sim$ 5\arcsec.

\subsection{Discussion}

Arising from the central region of IC~342, but being displaced from the nucleus, the H$_2$O maser is likely associated with a prominent star forming region.

\begin{figure}
\centering
\includegraphics[width=8cm]{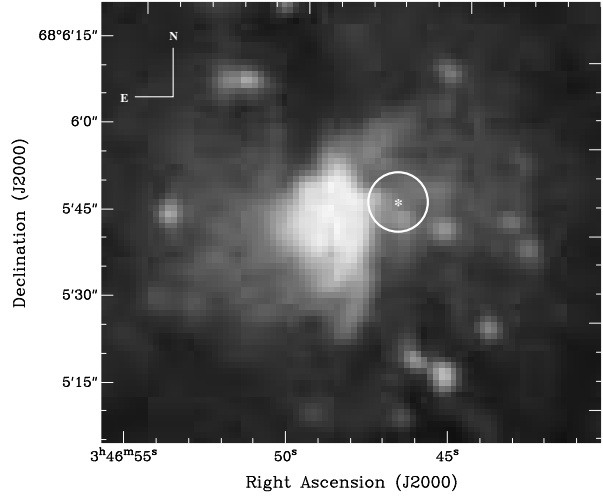}
\caption{XDSS B-band image of the central region of IC~342. The asterisk indicates the water maser emitting position. The radius of the circle indicates its positional error as outlined in Sect.\,3.2. \label{mappadss}}
\end{figure}

Fig.~\ref{mappadss} shows an optical B-band image of the central region of IC~342, taken from the XDSS\footnote{The Digitized Sky Surveys were produced at the Space Telescope Institute under U.S. Government grant NAGW-2166. The images of these surveys are based on photographic data obtained using the Oschin Schmidt Telescope on Palomar Mountain and the UK Schmidt Telescope. The plates were processed into the present compressed digital form with the permission of these institutions.}. The maser emitting region (white circle) coincides with an arc-like structure extending E-S and is associated with a chain of sources that appear to be HII regions. 

Occasionally, prominent galactic star forming regions show narrow ($\sim$0.5 km\,s$^{-1}$) flaring components that are exceptionally bright. Such flares were observed in W\,31\,A and W\,49 (\cite{lilje89}; \cite{lekht95}). The prototype of these flares is the 8\,km\,s$^{-1}$ super maser in Orion-KL (e.g.\ \cite{garay89}). During several months, the narrow highly linearly polarized maser reached flux densities in excess of 5\,MJy. Surpassing the flux of any other velocity component by more than an order of magnitude and reaching a peak luminosity of $L_{\rm H_2O}$ $\sim$ 10$^{-2}$\,L$_{\odot}$, the feature seems to be similar to that seen in IC~342. While luminosity, linewidth, and flux variations are reminiscent of the 8\,km\,s$^{-1}$ super maser in Orion-KL, our H$_2$O profiles show a significant velocity shift between April 22 and May 7 (Figs.~\ref{flare}b and c) that has not been seen in the Orion-KL flaring component. A simple model for explaining both the flare and the shift is, in our opinion, that the latter has a kinematic origin. Adopting a scenario of a chance alignment of two masing clouds along the line-of-sight, motion of the foreground relative to the background cloud along the plane of the sky and velocity structure in the foreground cloud could explain the observed data. A velocity gradient in the foreground cloud would then first shift the line velocity; once velocities are reached that are not matched within the background cloud, the flux density of the maser rapidly drops. Assuming that the distance between these clouds is $\ll$1.8\,Mpc and that their relative velocity in the plane of the sky is at the order of 100\,km\,s$^{-1}$, this implies a cloud velocity gradient of up to 1\,km\,s$^{-1}$/AU during the time the source was monitored.

An almost identical scenario was proposed by \cite{boboltz98} to account for the velocity shift of the flaring water maser component at --66\,km\,s$^{-1}$ in W49N. The similarity in velocity shift ($\sim$ 0.5\,km\,s$^{-1}$) and time scale (58 days) of the event in W49N w.r.t the one in IC~342, indicates a common origin.  

\section{Extragalactic maser detection rates}

While typical searches for extragalactic maser sources have only yielded detection rates between zero (e.g.\ \cite{henkel98}) and a few percent (e.g.\ \cite{henkel84}; \cite{braatz96}), there exists one sample with detection rates $>$10\%: These are the northern ($\delta$ $>$ --30$^{\circ}$) extragalactic IRAS (Infrared Astronomy Satellite) point sources with 100$\mu$m fluxes in excess of 50\,Jy (for a source list, see \cite{henkel86}; Maffei 2 with $S_{\rm 100\mu m}$ $\sim$ 200\,Jy should be added to the list). There is a total of 44 galaxies, two ultraluminous galaxies at intermediate distances (NGC\,3690 and Arp\,220) and 42 nearby sources ($V$ $<$ 3000\,km\,s$^{-1}$). Among these, seven (16\%) are known to contain H$_2$O masers in their nuclear region. Two of these contain megamasers (NGC~1068 and NGC~3079), two are possibly nuclear kilomasers (NGC~253 and M~51; for details, see Sect.\,1), and three are associated with prominent sites of star formation (IC~10, IC~342, M~82). Among the subsample of 19 sources with 100$\mu$m fluxes in excess of 100\,Jy, five were so far detected in H$_2$O, yielding a detection rate in excess of 20\%. Since few deep integrations have been obtained toward these sources, more H$_2$O detections can be expected from this promising sample in the near future.

\begin{acknowledgements}

We wish to thank Nikolaus Neininger for useful discussion. We are also endebted to the operators at the 100-m telescope, and to Michael Rupen and the NRAO analysts, for their cheerful assistance with the observations.

\end{acknowledgements}

%
%
\end{document}